\documentclass[pdflatex,iicol,sn-mathphys-num]{sn-jnl}

\usepackage{graphicx}
\usepackage{multirow}
\usepackage{amsmath,amssymb,amsfonts}
\usepackage{amsthm}
\usepackage{mathrsfs}
\usepackage[title]{appendix}
\usepackage{xcolor}
\usepackage{textcomp}
\usepackage{manyfoot}
\usepackage{booktabs}
\usepackage{algorithm}
\usepackage{algorithmicx}
\usepackage{algpseudocode}
\usepackage{listings}

\usepackage{physics}

\usepackage{color}

\usepackage{ulem}      
\usepackage{caption}   
\usepackage{framed}    

\newcommand{\vkap}{\vb*{\kappa}}  

\theoremstyle{thmstyleone}

\theoremstyle{thmstyletwo}

\theoremstyle{thmstylethree}

\begin{document}

\title{Phenomenological refinement of \textit{p}-\textit{d} elastic scattering descriptions towards the 3NF study in nuclei via the (\textit{p},\textit{pd}) reaction}

\author*[1]{Yoshiki Chazono}
\email{chazono@rcnp.osaka-u.ac.jp}
\author[2,3]{Tokuro Fukui}
\author[1,3,4]{Futoshi Minato}
\author[1,5]{Yukinobu Watanabe}
\author[1]{Kazuyuki Ogata}

\affil[1]{Department of Physics, Kyushu University, Fukuoka 819-0395, Japan}
\affil[2]{Faculty of Arts and Science, Kyushu University, Fukuoka, 819-0395, Japan}
\affil[3]{RIKEN Nishina Center for Accelerator-Based Science, Wako, Saitama 351-0198, Japan}
\affil[4]{Nuclear Data Center, Japan Atomic Energy Agency, Tokai, Ibaraki 319-1195, Japan}
\affil[5]{Department of Advanced Energy Science and Engineering, Kyushu University, Fukuoka 816-8580, Japan}


\maketitle

\section*{Abstract}
The proton-induced deuteron knockout ($p,pd$) reaction is expected to be a powerful tool for probing three-nucleon forces (3NFs) in nuclear medium since it can be essentially regarded as the $p$-$d$ elastic scattering inside nuclei.
One of the important points in the theoretical description of the ($p,pd$) reaction is to calculate the $p$-$d$ scattering in a nucleus quantitatively using effective interactions.
This work aims to develop a phenomenological approach to improve the quantitativity of the $p$-$d$ scattering cross section in free space calculated with effective interactions.
The $p$-$d$ elastic amplitude is decomposed into a 2N part, described using 2N effective interactions, and a residual part, which the 2N part cannot describe.
The latter is approximated by a superposition of Legendre polynomials, with coefficients treated as adjustable parameters.
These parameters are determined to reproduce experimental $p$-$d$ differential cross-section data at various incident energies.
The obtained parameters exhibit smooth energy dependence, which is approximated by quadratic functions.
The numerical results with the analytic energy dependence also reproduce the experimental data.
The developed approach works well for improving the $p$-$d$ scattering cross section in a wide range of incident energies.
This work can be regarded as the first step toward the description of ($p,pd$) reactions taking 3NF effect in nuclear medium into account.

\section{Introduction\label{sec:introduction}}
Recent progress in chiral effective field theory~\cite{Weinberg1979327, Epelbaum2006654, MACHLEIDT20111} has refined our knowledge of three-nucleon forces (3NFs) in many-nucleon systems.
For instance, the chiral 3NF~\cite{WEINBERG1992114, PhysRevC.49.2932, PhysRevLett.85.2905} plays an essential role in the description of $^{10}\mathrm{B}$~\cite{PhysRevLett.99.042501}, understanding the spin-orbit splitting of light nuclei~\cite{FUKUI2024138839}, the explanation of the limit of oxygen isotopes~\cite{PhysRevLett.105.032501}, and the realistic prediction of nuclear matter saturation~\cite{10.3389/fphy.2019.00213}.

One of the typical approaches to study the 3NF in medium is based on an approximation that the 3NF is expressed as a density dependent two-nucleon (2N) effective interaction.
Because the proton-induced proton knockout ($p,2p$) reactions are regarded as proton-proton ($p$-$p$) elastic scattering inside nuclei, they have been utilized to clarify the in-medium modification to the $p$-$p$ interaction~\cite{TWakasa17}.
Quite recently, Ref.~\cite{KMinomo17a} revealed that the triple differential cross section of the $^{40}\mathrm{Ca}(p,2p)^{39}\mathrm{K}$ reaction, for which the $0p_{3/2}$ orbit proton is knocked out, is sensitive to the chiral 3NF effect on the $p$-$p$ cross section in nuclear medium.

As an extension of Ref.~\cite{KMinomo17a}, one may consider using the deuteron knockout ($p,pd$) reaction to investigate the 3NF effect on the $p$-$d$ elastic cross section in nuclei.
The ($p,pd$) reaction has attracted attention as a means to probe the presence of deuterons in nuclei~\cite{TUesaka24}.
Theoretically, one of the authors (Y.~C.) and collaborators developed a novel reaction model, CDCCIA, for describing ($p,pd$) with explicitly accounting for the fragility of the deuteron~\cite{YChazono22}.
One of the key ingredients of CDCCIA is the transition matrix for the $p$-$d$ reaction system, which is formulated in terms of a 2N effective interaction.
This approach is suitable for the aforementioned strategy of investigating 3NF effect through the density dependence of the 2N effective interaction.

To lay the groundwork for 3NF studies using ($p,pd$) reactions, it is necessary to have reliable $p$-$d$ elastic cross sections in free space over a range of energies and scattering angles.
We plan to explore in-medium 3NF effects via the density dependence of the 2N effective interaction.
Therefore, a model for the $p$-$d$ elastic cross sections in free space based on an effective interaction is needed.
However, as shown in Ref.~\cite{YChazono22}, calculations using the Franey-Love 2N effective interaction~\cite{MAFraney85} tend to underestimate the experimental cross sections around the minima.
This underestimation has also been reported in Faddeev calculations that include only the 2N force~\cite{HWitala98, HWitala22}.
In our approach, we resolve this discrepancy by adding a correction amplitude to the $p$-$d$ transition matrix used in Ref.~\cite{YChazono22}, which might include effects beyond the 2N effective interaction.
This additional amplitude is parameterized as a superposition of Legendre polynomials, and the energy dependence of each parameter is described by an analytic function.

This paper is organized as follows.
In Sec.~\ref{sec:framework}, we explain our approach for improving the differential cross section of the $p$-$d$ scattering phenomenologically.
In Sec.~\ref{sec:result}, the inputs used in the numerical calculation are mentioned.
We also show the obtained results compared with the experimental data and discuss how well our approach works.
Finally, the summary and perspective are given in Sec.~\ref{sec:summary}.

\section{Theoretical framework\label{sec:framework}}
We consider the differential cross section of the $p$-$d$ elastic scattering.
All quantities are evaluated in the center-of-mass (c.m.) frame of the $p$-$d$ system, and those with the superscript L are evaluated in the laboratory (L) frame.
We adopt the isospin representation, i.e., the proton and neutron are distinguished by the third component of their isospin, in numerical calculations.

We describe a $p$-$d$ elastic scattering cross section as
\begin{align}
\dv{\sigma}{\Omega} =
C_\textrm{corr}
\qty(\frac{\mathcal{M}_{pd}}{2 \pi \hbar^2})^2
\frac{1}{6}
\abs{\tilde{t}_\textrm{2N} + \tilde{t}_\textrm{res}}^2,
\label{eq:CS_pd}
\end{align}
where $\mathcal{M}_{pd}$ is the reduced energy of the $p$-$d$ system.
The $p$-$d$ elastic amplitude is decomposed into the two parts, $\tilde{t}_\textrm{2N}$ and $\tilde{t}_\textrm{res}$.
The former is the contribution from 2N effective interactions, and the latter represents the residual part that $\tilde{t}_\textrm{2N}$ cannot describe in the $p$-$d$ elastic scattering process.
$C_\textrm{corr}$ is the phenomenological correction factor to reproduce the $\dv*{\sigma}{\Omega}$ at forward angles~\cite{YChazono22}.
Note that the value of $C_\textrm{corr}$ is slightly different from that in Ref.~\cite{YChazono22}, where only $\tilde{t}_\textrm{2N}$ was considered, because we have determined it at the most forward angle at each energy to exclude contributions of $\tilde{t}_\textrm{res}$ as much as possible.
Those values are listed in Table~\ref{tab:parameter}.

In the same manner as used in Ref.~\cite{YChazono22}, we calculate $\tilde{t}_\textrm{2N}$ using 2N effective interactions as
\begin{align}
\tilde{t}_\textrm{2N} =
\mel**{\Phi_d, \vkap'
}{\qty[t_{pp} + t_{pn}] \hat{\mathcal{A}}
}{\Phi_d, \vkap},
\label{eq:t_2N}
\end{align}
where $\Phi_d$ is the antisymmetrized ground-state wave function of the deuteron.
The vectors $\vkap$ and $\vkap'$ are the relative momenta (in units of $\hbar$) between the incident proton and the c.m. of the deuteron in the initial and final channels, respectively.
$t_{pp}$ and $t_{pn}$ are the proton-proton and proton-neutron effective interactions in free space, respectively, which are essentially the same since we adopt the isospin representation.
These are defined as
\begin{align}
t_\textrm{NN} \ket{\vb*{k}} = V_\textrm{NN} \ket{\chi_{\vb*{k}}} \quad
(\textrm{N} = p \ \textrm{or} \ n),
\end{align}
where $\vb*{k}$ is the relative momentum (in units of $\hbar$) between two nucleons.
$V_\textrm{NN}$ is a bare or realistic nuclear force and $\chi_{\vb*{k}}$ is the solution of the Schr\"{o}dinger equation of the two-nucleon system with $V_\textrm{NN}$.
The antisymmetrization operator $\hat{\mathcal{A}}$ is defined by
\begin{align}
\hat{\mathcal{A}} \equiv
\frac{1}{\sqrt{3}} \qty[1 - \hat{P}_{pp} - \hat{P}_{pn}],
\label{eq:AntiSymm}
\end{align}
where $\hat{P}_{pp} \ (\hat{P}_{pn})$ exchanges the incident proton and the proton (neutron) in the deuteron.
With the use of the antisymmetrized deuteron wave function, the antisymmetrization of the 3N system requires the normalization coefficient $1 / \sqrt{3}$ in Eq.~\eqref{eq:AntiSymm}.

We assume $\tilde{t}_\textrm{res}$ is represented as a superposition of Legendre polynomials $P_l$, i.e.,
\begin{align}
\tilde{t}_\textrm{res} =
\sum_{l = 0}^{l_\textrm{max}}
C_l P_l (\cos{\theta}),
\label{eq:t_res}
\end{align}
where $l$ is the angular momentum transfer (in units of $\hbar$) and $\theta$ is the scattering angle.
The complex coefficients $C_l$ are determined phenomenologically so that Eq.~\eqref{eq:CS_pd} reproduces the experimental data.

For convenience, we also define the cross sections including only $\tilde{t}_\textrm{2N}$ and $\tilde{t}_\textrm{res}$ as
\begin{align}
\qty(\dv{\sigma}{\Omega})_\textrm{2N} =
C_\textrm{corr}
\qty(\frac{\mathcal{M}_{pd}}{2 \pi \hbar^2})^2
\frac{1}{6}
\abs{\tilde{t}_\textrm{2N}}^2
\label{eq:CS_pd_2N}
\end{align}
and
\begin{align}
\qty(\dv{\sigma}{\Omega})_\textrm{res} =
C_\textrm{corr}
\qty(\frac{\mathcal{M}_{pd}}{2 \pi \hbar^2})^2
\frac{1}{6}
\abs{\tilde{t}_\textrm{res}}^2,
\label{eq:CS_pd_res}
\end{align}
respectively.
The former represents the result obtained using only the 2N effective interaction, which is identical to the cross section shown in Ref.~\cite{YChazono22}.  
The latter corresponds to the contribution from the newly added amplitude of Eq.~\eqref{eq:t_res}; for a simple notation, we refer to it as the $p$-$d$ cross section including only the residual part.
The factors $C_\textrm{corr}$ in these equations are the same in Eq.~\eqref{eq:CS_pd}.
Since there is interference between $\tilde{t}_\textrm{2N}$ and $\tilde{t}_\textrm{res}$, the sum of Eqs.~\eqref{eq:CS_pd_2N} and \eqref{eq:CS_pd_res} does not necessarily agree with the cross section of Eq.~\eqref{eq:CS_pd}.

\section{Result and discussion\label{sec:result}}

\subsection{Input\label{sec:input}}
As one of the indicators for determining the coefficients $C_l$ in Eq.~\eqref{eq:t_res}, we used $\chi^2$ defined by
\begin{align}
\chi^2 \equiv
\frac{1}{N} \sum_{i = 1}^N \frac{1}{\epsilon^2 (\theta_i)}
\qty[\qty(\dv{\sigma}{\Omega})_\textrm{exp}^{(i)}
- \qty(\dv{\sigma}{\Omega})_\textrm{theor}^{(i)}]^2
\label{eq:chisq}
\end{align}
at proton-incident energy $T_p^\textrm{L}$.
Label $i$ specifies the scattering angle where an experimental cross section is measured, and $N$ is the number of data at each incident energy.
An experimental error at each scattering angle is represented by $\epsilon(\theta_{i})$.
We minimized Eq.~\eqref{eq:chisq} with the Marquardt method~\cite{KLevenberg44, DWMarquardt63} using multiple randomly generated initial values of $C_l$, and then selected the $C_l$ values using $\chi^2$ and smoothness of $T_p^\textrm{L}$ dependence as indicators following Ref.~\cite{AJKoning03}.

We carried out the above parameter search at eight incident energies: $108$, $120$, $135$, $150$, $155$, $170$, $190$, and $250$~MeV, with $l_\textrm{max}$ set to 1.
The experimental cross sections used for the search were taken from Ref.~\cite{KErmisch05} ($108$, $120$, $135$, $150$, $170$, and $190$~MeV), Ref.~\cite{KKuroda64} ($155$~MeV), and Ref.~\cite{KHatanaka02} ($250$~MeV); the experimental statistical errors are less than $1\%$ and the systematic ones are about $5\%$.
One of the purposes of this paper is to explore $C_l$ having a relatively smooth incident-energy dependence, not to optimize them at each energy.
Hence, we set $\epsilon$ to $5\%$ similar to the systematic errors.

The 2N effective interaction based on the Bonn-B potential~\cite{RMachleidtP89}, which was parameterized by the Melbourne group~\cite{KAmosB00}, was used for $t_{pp}$ and $t_{pn}$ in Eq.~\eqref{eq:t_2N}.
Because the Coulomb interaction between $p$ and $d$ is not included in our calculation, we used only the data at angles larger than $20^\circ$.

\subsection{Parameter\label{sec:parameter}}

\begin{figure}[htpb]
 \centering
 \includegraphics[width=0.50\textwidth,bb=0 0 648 576]{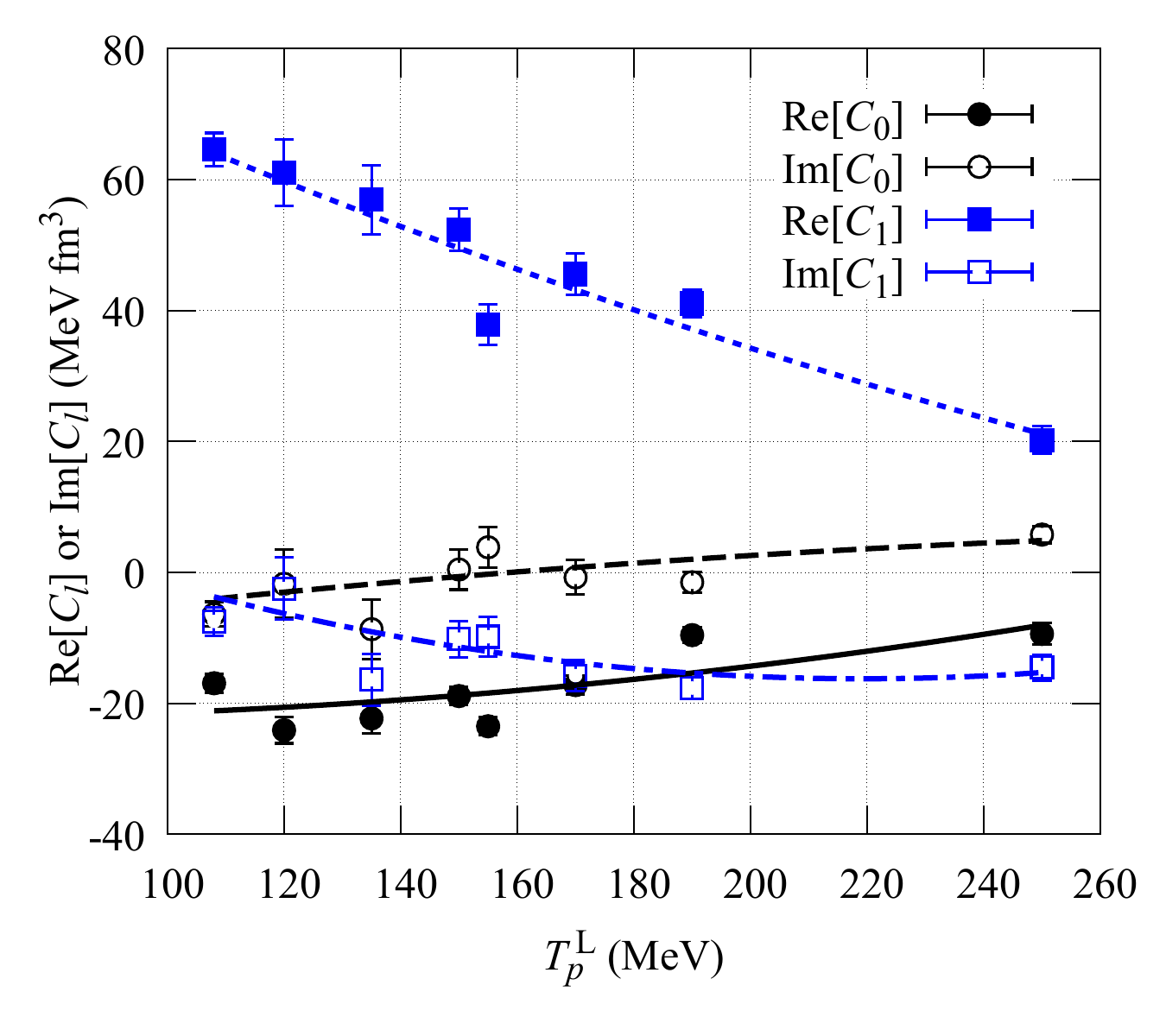}
 \caption{Incident-energy dependence of $C_l$.
 The filled and open circles denote the real and imaginary parts of $C_0$, respectively, while the square symbols correspond to the real and imaginary parts of $C_1$.
 The approximated behaviors of the filled (open) circles and squares, Eq.~\eqref{eq:C_L}, are plotted by the solid and dotted (dashed and dot-dashed) lines, respectively.
 \label{fig:parameter}}
\end{figure}

\begin{table*}[htbp]
 \centering
 \caption{The determined coefficients $C_0$ and $C_1$ of Eq.~\eqref{eq:t_res}, and the values of $C_\textrm{corr}$.
 \label{tab:parameter}}
 \begin{tabular}{crrrrc}
 \toprule
 $T_p^\textrm{L}$ & \multicolumn{2}{c}{$C_0$ (MeV fm$^3$)} & \multicolumn{2}{c}{$C_1$ (MeV fm$^3$)} & $C_\textrm{corr}$ \\
 (MeV) & \multicolumn{1}{c}{$\mathrm{Re}$} & \multicolumn{1}{c}{$\mathrm{Im}$}
       & \multicolumn{1}{c}{$\mathrm{Re}$} & \multicolumn{1}{c}{$\mathrm{Im}$} \\
 \midrule
 $108$ & $-1.692\mathrm{E+}1$ & $-6.379\mathrm{E+}0$ & $+6.462\mathrm{E+}1$ & $-7.481\mathrm{E+}0$ & $0.792$ \\
 $120$ & $-2.410\mathrm{E+}1$ & $-1.735\mathrm{E+}0$ & $+6.101\mathrm{E+}1$ & $-2.452\mathrm{E+}0$ & $0.749$ \\
 $135$ & $-2.229\mathrm{E+}1$ & $-8.665\mathrm{E+}0$ & $+5.693\mathrm{E+}1$ & $-1.640\mathrm{E+}1$ & $0.818$ \\
 $150$ & $-1.888\mathrm{E+}1$ & $+4.531\mathrm{E-}1$ & $+5.235\mathrm{E+}1$ & $-1.018\mathrm{E+}1$ & $0.701$ \\
 $155$ & $-2.349\mathrm{E+}1$ & $+3.848\mathrm{E+}0$ & $+3.785\mathrm{E+}1$ & $-9.815\mathrm{E+}0$ & $0.752$ \\
 $170$ & $-1.720\mathrm{E+}1$ & $-7.544\mathrm{E-}1$ & $+4.557\mathrm{E+}1$ & $-1.576\mathrm{E+}1$ & $0.721$ \\
 $190$ & $-9.573\mathrm{E+}0$ & $-1.498\mathrm{E+}0$ & $+4.109\mathrm{E+}1$ & $-1.765\mathrm{E+}1$ & $0.794$ \\
 $250$ & $-9.392\mathrm{E+}0$ & $+5.787\mathrm{E+}0$ & $+2.024\mathrm{E+}1$ & $-1.452\mathrm{E+}1$ & $0.711$ \\
 \bottomrule
 \end{tabular}
\end{table*}

We show the proton incident-energy dependence of the obtained parameters in Fig.~\ref{fig:parameter}.
The real and imaginary parts of $C_0$ are represented by the filled and open circles, respectively, while those of $C_1$ are denoted by the square symbols.
The error bars indicate the fitting uncertainties.
The values of these parameters are listed in Table~\ref{tab:parameter}.
One can find that both $C_0$ and $C_1$ have the smooth $T_p^\textrm{L}$ dependence.
Quadratic functions are then used to approximate their energy dependence:
\begin{align}
C_l \approx \alpha_l (T_p^\textrm{L})^2 + \beta_l T_p^\textrm{L} + \gamma_l \quad
(l = 0 \ \textrm{and} \ 1).
\label{eq:C_L}
\end{align}
The coefficients $\alpha_l$, $\beta_l$, and $\gamma_l$ are summarized in Table~\ref{tab:coefficient}.
The approximated $T_p^\textrm{L}$ dependence of $C_l$ expressed by Eq.~\eqref{eq:C_L} are also plotted in Fig.~\ref{fig:parameter} by the solid ($\textrm{Re}[C_0]$), dashed ($\textrm{Im}[C_0]$), dotted ($\textrm{Re}[C_1]$), and dot-dashed ($\textrm{Im}[C_1]$) lines.
Note that we excluded the $C_l$ values at $135$~MeV when determining $\alpha_l$, $\beta_l$, and $\gamma_l$ to test the predictivity of Eq.~\eqref{eq:C_L} in Sec.~\ref{sec:crosssec}.
As shown later in Figs.~\ref{fig:CS_pd_1} and \ref{fig:CS_pd_2}, the differential cross sections of $p$-$d$ scattering calculated using Eq.~\eqref{eq:C_L} reproduce the experimental data with minor uncertainties.
Therefore, Eq.~\eqref{eq:C_L} can be used generally to describe $\tilde{t}_\textrm{res}$ of Eq.~\eqref{eq:t_res} at energies between $100$ and $250$~MeV.

\begin{table*}[htbp]
 \centering
 \caption{Coefficients $\alpha_l$, $\beta_l$, and $\gamma_l$ of Eq.~\eqref{eq:C_L}.
 \label{tab:coefficient}}
 \begin{tabular}{crrrrrrrrrrrr}
 \toprule
     & \multicolumn{2}{c}{$\alpha_l$ (MeV$^{-1}$ fm$^3$)}
     & \multicolumn{2}{c}{$\beta_l$ (fm$^3$)}
     & \multicolumn{2}{c}{$\gamma_l$ (MeV fm$^3$)} \\
 $l$ & \multicolumn{1}{c}{$\mathrm{Re}$} & \multicolumn{1}{c}{$\mathrm{Im}$}
     & \multicolumn{1}{c}{$\mathrm{Re}$} & \multicolumn{1}{c}{$\mathrm{Im}$}
     & \multicolumn{1}{c}{$\mathrm{Re}$} & \multicolumn{1}{c}{$\mathrm{Im}$} \\
 \midrule
 $0$ & $+3.610\mathrm{E-}4$ & $-1.862\mathrm{E-}4$ & $-3.708\mathrm{E-}2$ & $+1.295\mathrm{E-}1$ & $-2.134\mathrm{E+}1$ & $-1.587\mathrm{E+}1$ \\
 $1$ & $+4.307\mathrm{E-}4$ & $+1.012\mathrm{E-}3$ & $-4.561\mathrm{E-}1$ & $-4.436\mathrm{E-}1$ & $+1.083\mathrm{E+}2$ & $+3.239\mathrm{E+}1$ \\
 \bottomrule
 \end{tabular}
\end{table*}

\subsection{Cross section\label{sec:crosssec}}

\begin{figure*}[htbp]
 \includegraphics[width=0.50\textwidth,bb=0 0 648 576]{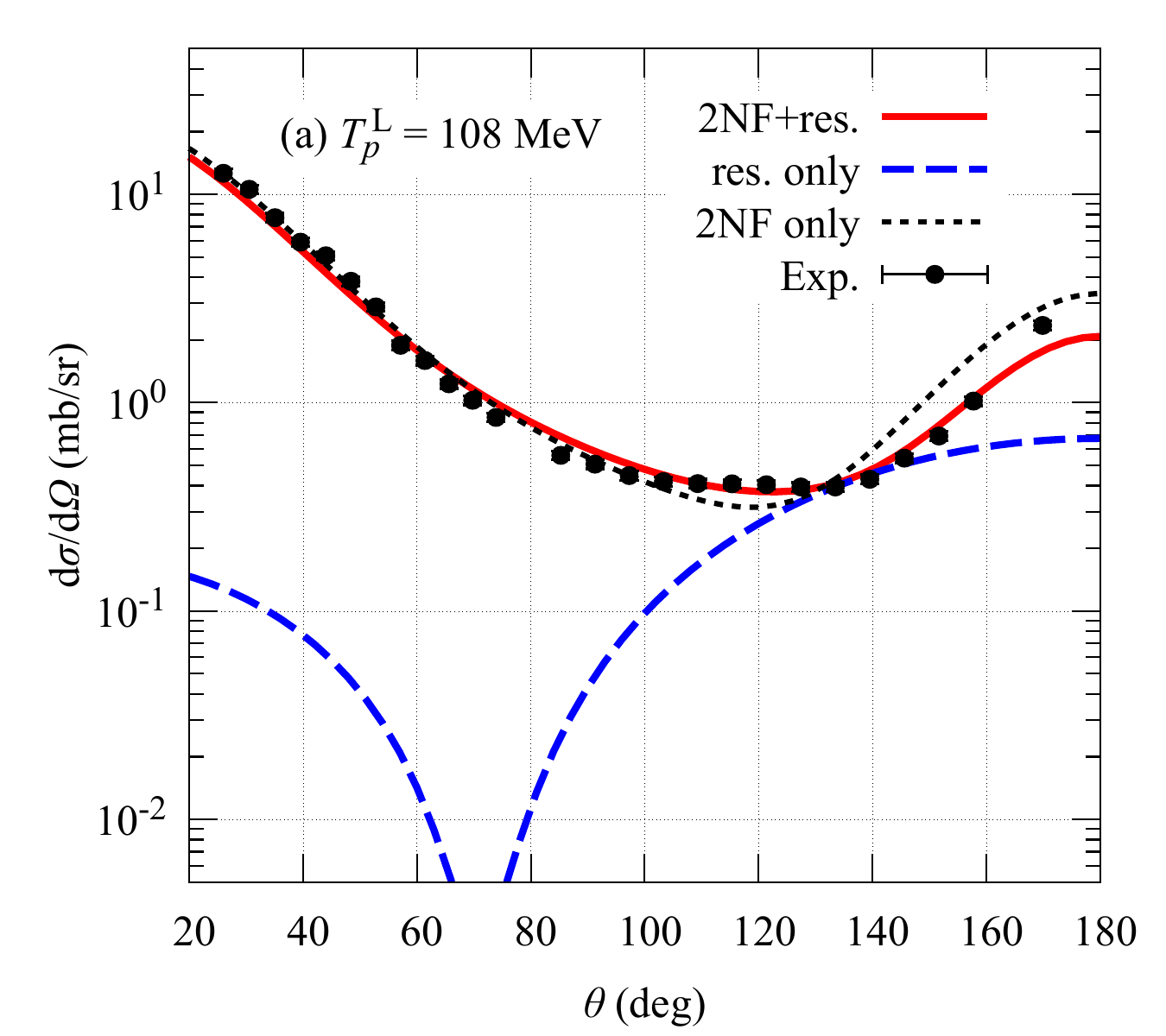}
 \includegraphics[width=0.50\textwidth,bb=0 0 648 576]{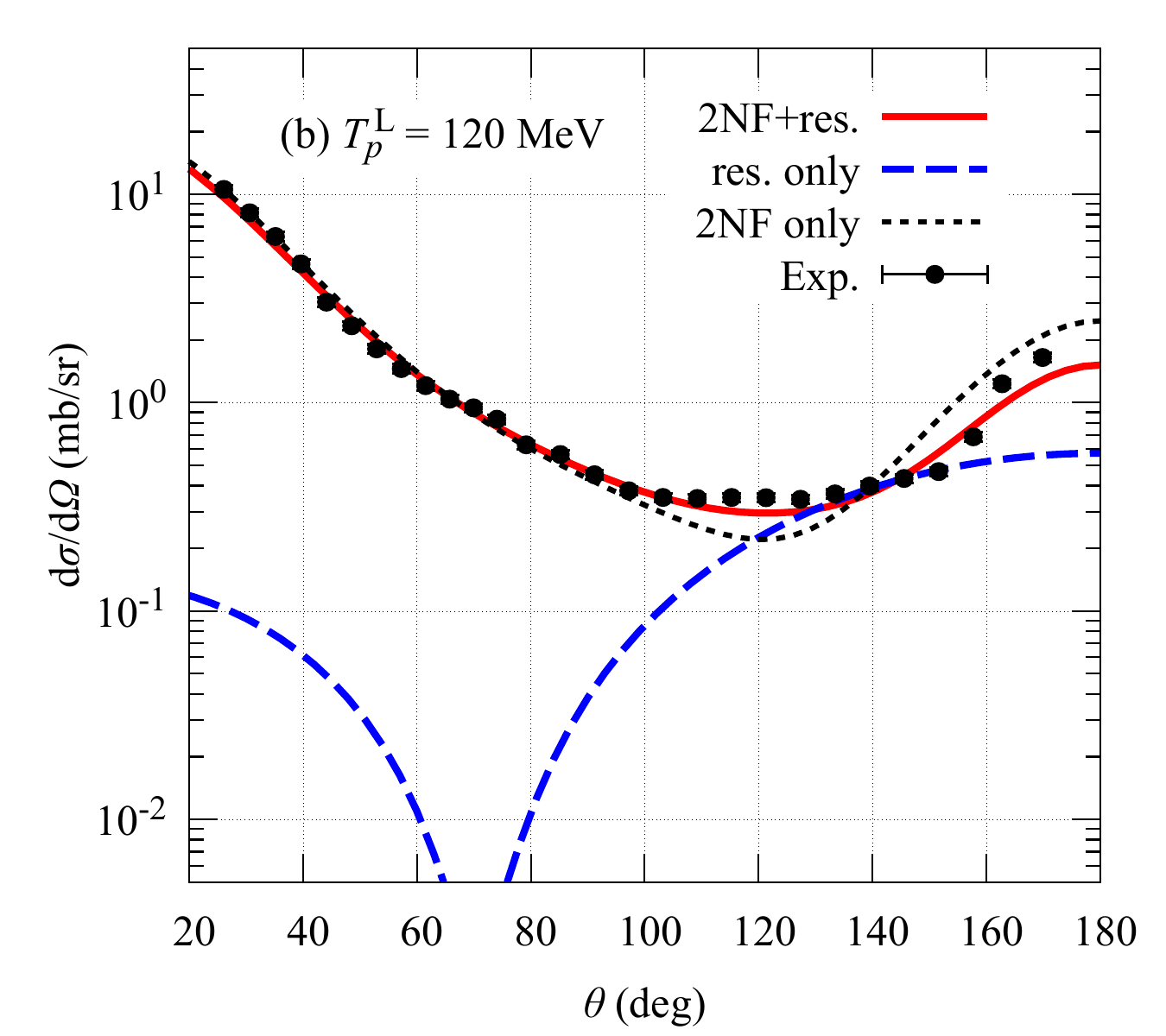}\\
 \includegraphics[width=0.50\textwidth,bb=0 0 648 576]{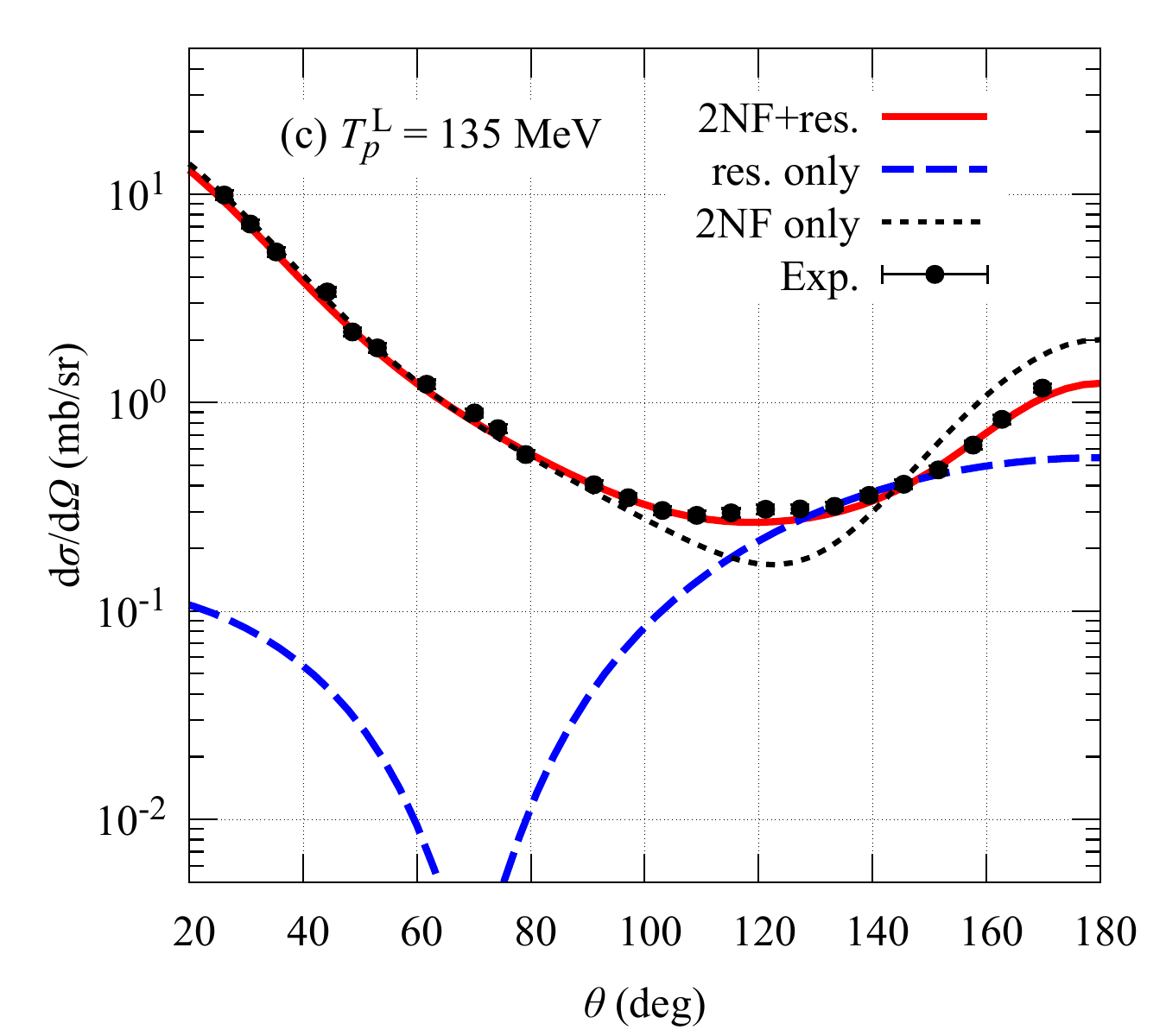}
 \includegraphics[width=0.50\textwidth,bb=0 0 648 576]{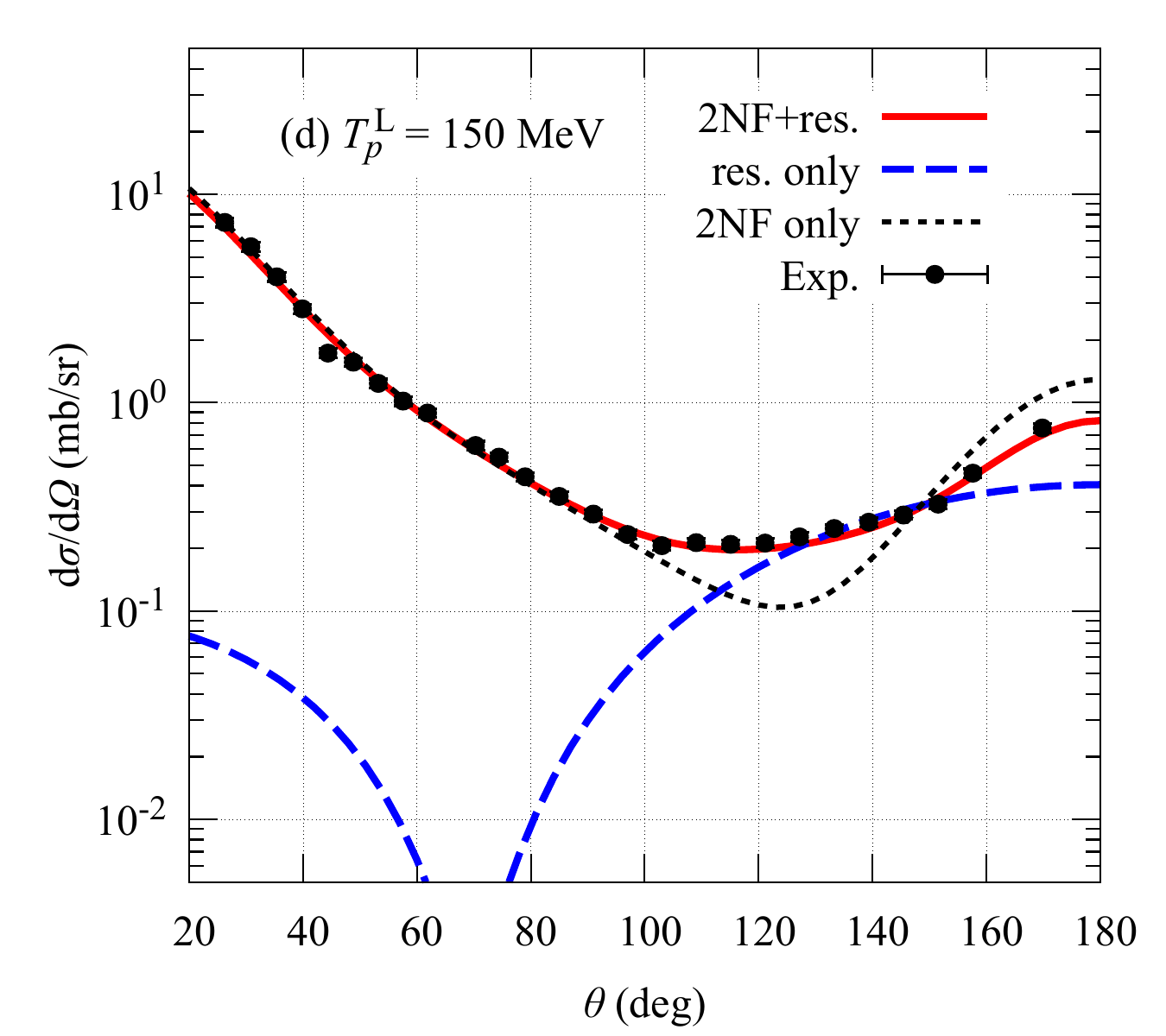}
 \caption{Differential cross sections of the $p$-$d$ elastic scattering  at $T_p^\textrm{L} =$ (a) $108$, (b) $120$, (c) $135$, and (d) $150$~MeV as a function of the c.m. scattering angle of the system.
  The dots are the experimental data taken from Ref.~\cite{KErmisch05}.
  The solid lines are the numerical results of Eq.~\eqref{eq:CS_pd} with the parameters evaluated by Eq.~\eqref{eq:C_L}, while the dotted and dashed ones are the results obtained using Eqs.~\eqref{eq:CS_pd_2N} and \eqref{eq:CS_pd_res}, respectively.
 \label{fig:CS_pd_1}}
\end{figure*}

\begin{figure*}[htbp]
 \includegraphics[width=0.50\textwidth,bb=0 0 648 576]{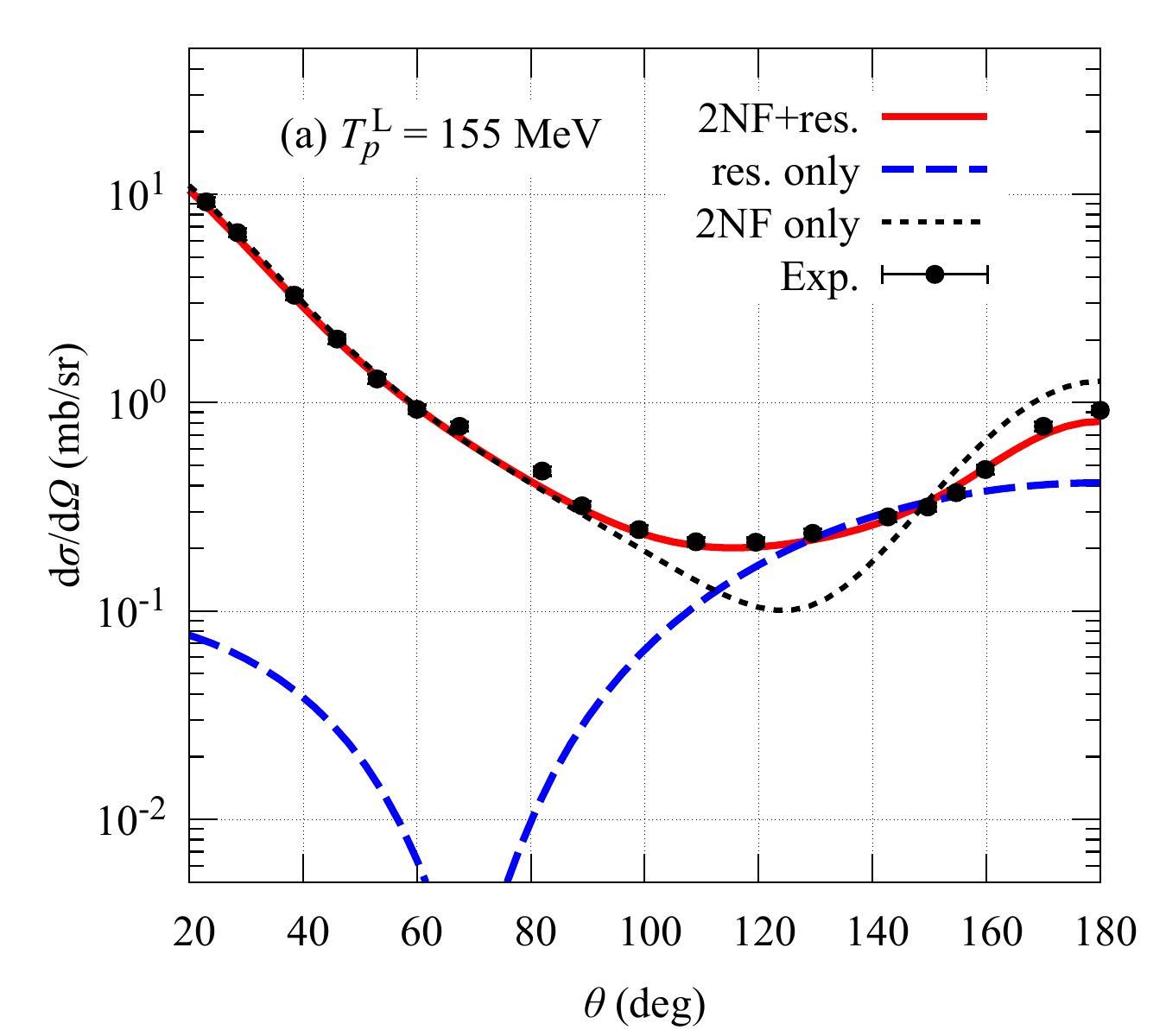}
 \includegraphics[width=0.50\textwidth,bb=0 0 648 576]{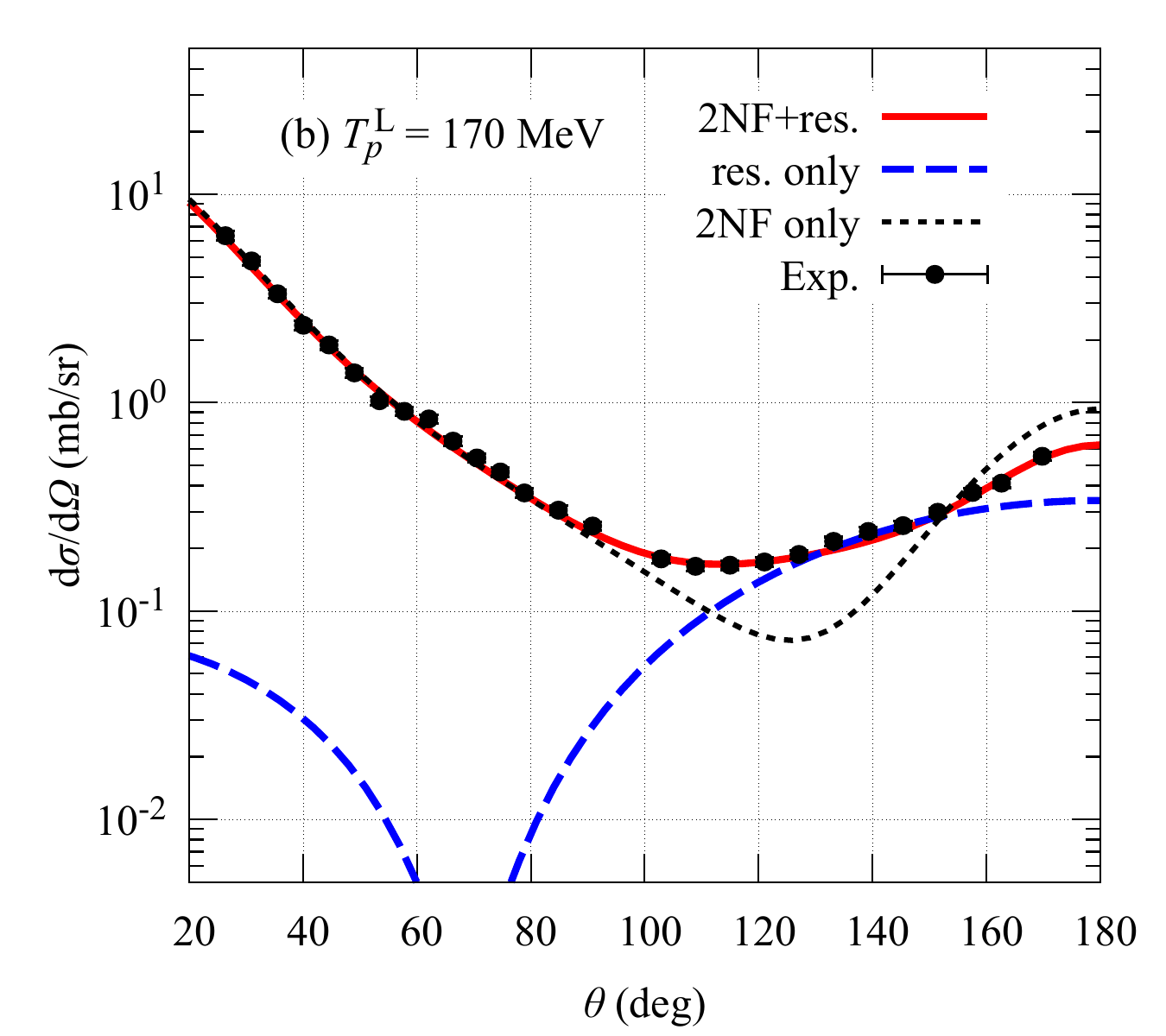}\\
 \includegraphics[width=0.50\textwidth,bb=0 0 648 576]{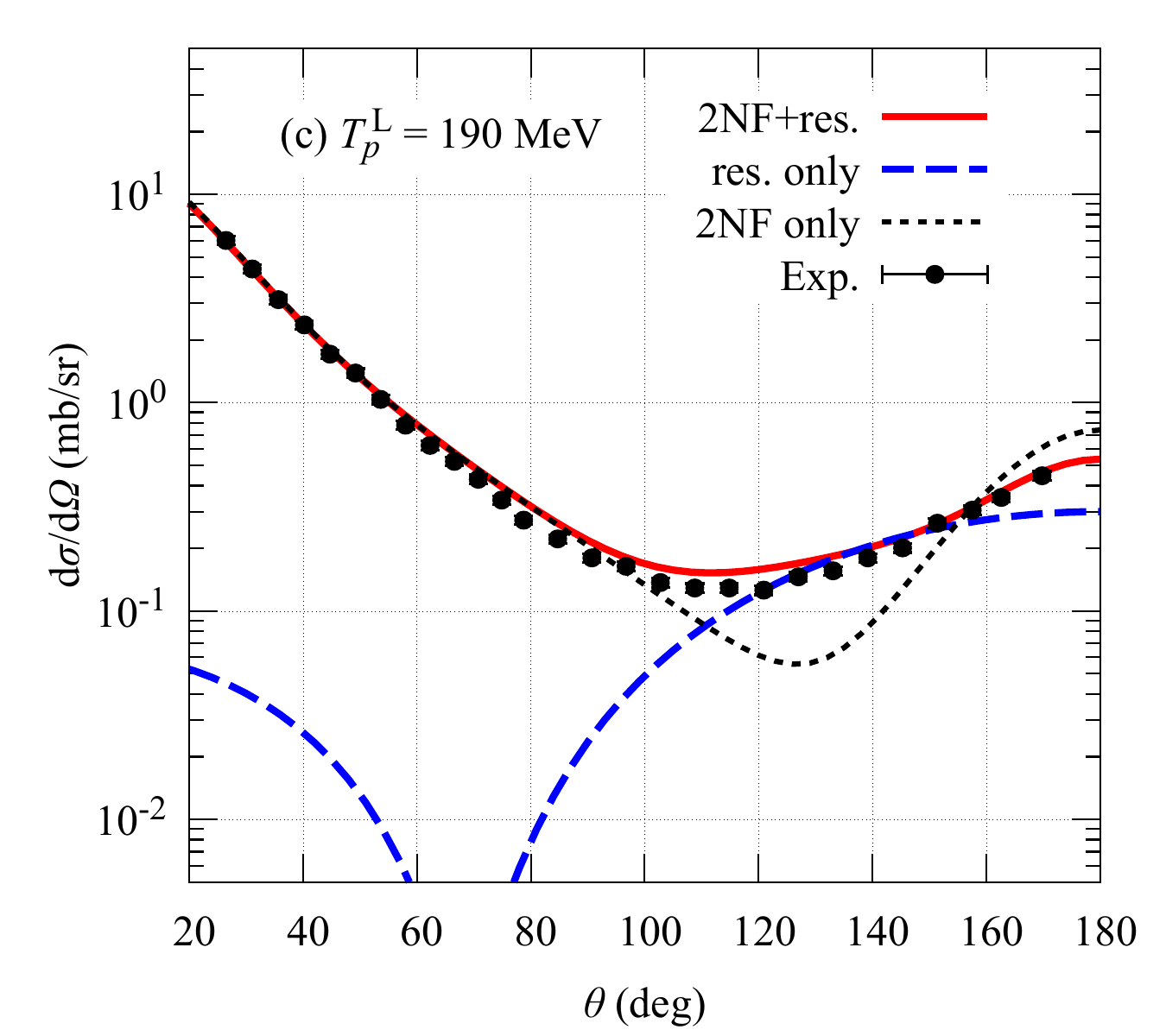}
 \includegraphics[width=0.50\textwidth,bb=0 0 648 576]{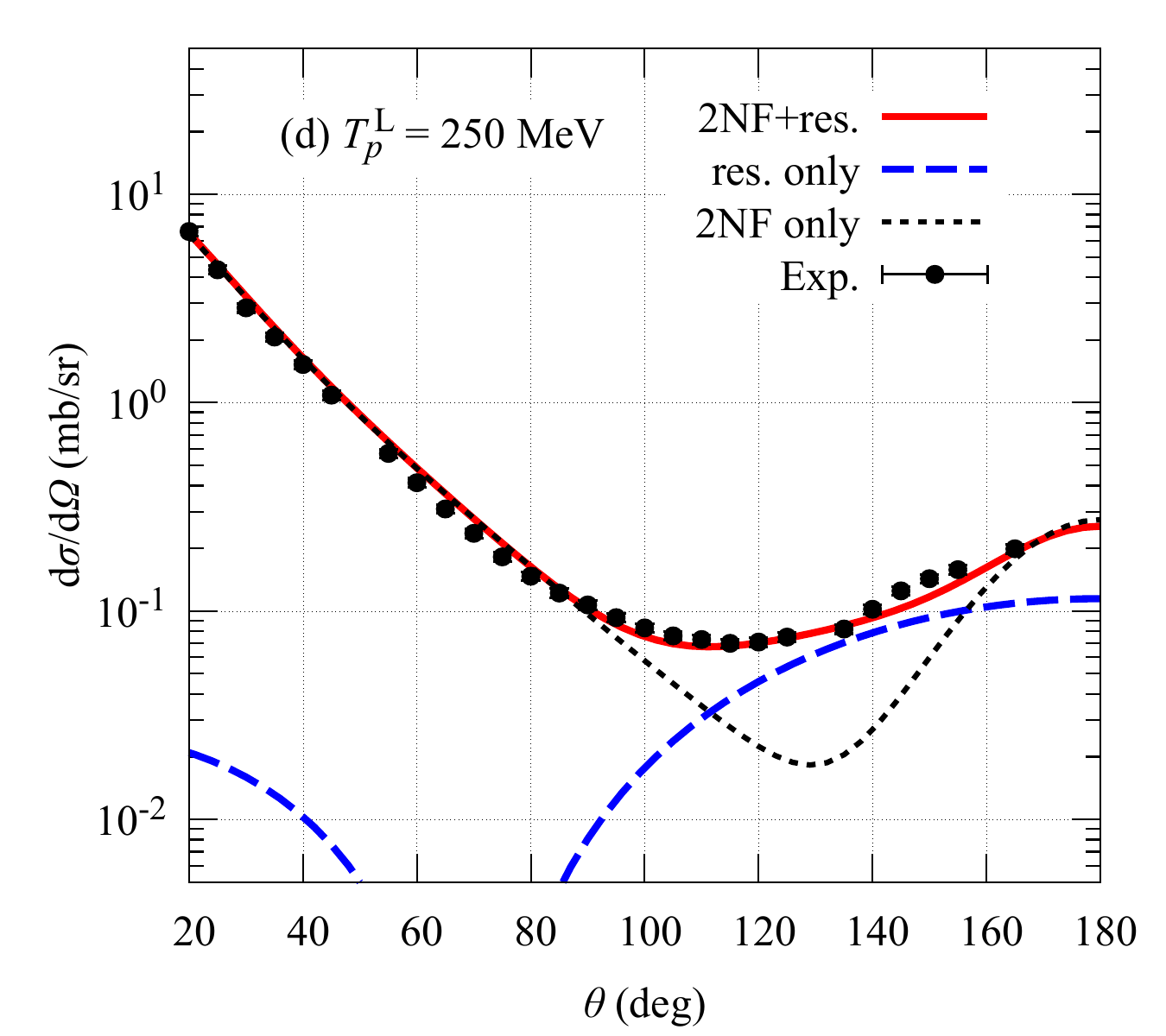}
 \caption{Same as Fig.~\ref{fig:CS_pd_1} but at $T_p^\textrm{L} =$ (a) $155$, (b) $170$, (c) $190$, and (d) $250$~MeV.
 The experimental data are taken from Refs.~\cite{KKuroda64} ($155$~MeV), \cite{KErmisch05} ($170$ and $190$~MeV), and \cite{KHatanaka02} ($250$~MeV).
 \label{fig:CS_pd_2}}
\end{figure*}

We show the $p$-$d$ elastic differential cross sections at $T_p^\textrm{L} =$ (a) $108$, (b) $120$, (c) $135$, and (d) $150$~MeV in Fig.~\ref{fig:CS_pd_1} as a function of the scattering angle $\theta$.
Figure~\ref{fig:CS_pd_2} shows the same as Fig.~\ref{fig:CS_pd_1} but at $T_p^\textrm{L} =$ (a) $155$, (b) $170$, (c) $190$, and (d) $250$~MeV.
In each panel, the dots denote the experimental data taken from Refs.~\cite{KErmisch05, KKuroda64, KHatanaka02}.

The dotted lines are the cross sections calculated using Eq.~\eqref{eq:CS_pd_2N}, i.e., only the 2N effective interaction is taken into account.
At all energies, the lines reproduce the experimental data at $\theta \lesssim 90^\circ$, but there are differences between the lines and dots at the larger angles.
In particular, the dotted lines obviously underestimate the data between $100^\circ$ and $150^\circ$.

The solid lines represent the results of Eq.~\eqref{eq:CS_pd} using the parameter evaluated by Eq.~\eqref{eq:C_L}.
The lines generally describe well the data at all angles in the energy range considered, including $135$~MeV.
Therefore, the parameters approximated by Eq.~\eqref{eq:C_L} can describe reasonably the energy dependence of $\tilde{t}_\textrm{res}$ in the proton incident-energy range from about $100$~MeV to $250$~MeV.
The uncertainty arising from those of the parameter search is almost $2\%$, comparable to the line width.
Comparing the solid and dotted lines, the inclusion of $\tilde{t}_\textrm{res}$ improves the angular dependence at $\theta \gtrsim 90^\circ$.
The role of $\tilde{t}_\textrm{res}$ appears to be qualitatively consistent with the 3NF contributions reported in Refs.~\cite{HWitala98, HWitala22}.
Although $\tilde{t}_\textrm{res}$ includes various contributions beyond 2N effective interactions, it may primarily reflect the effects of 3NFs.

The dashed lines are the numerical results of Eq.~\eqref{eq:CS_pd_res}.
As expected from the smooth energy dependence of $C_l$ shown in Fig.~\ref{fig:parameter}, the lines change gradually as the incident energy varies.
At forward angles up to around $90^\circ$, one sees that the contribution of 2N effective interaction is dominant.
In contrast, the contributions of $\tilde{t}_\textrm{2N}$ and $\tilde{t}_\textrm{res}$ to the $p$-$d$ elastic cross section are comparable.
From about $160^\circ$, the dotted lines are larger than the experimental data, but the solid ones reproduce the data.
This indicates that $\tilde{t}_\textrm{2N}$ and $\tilde{t}_\textrm{res}$ have opposite phases of each other and interfere destructively in the $p$-$d$ elastic cross sections.

\section{Summary\label{sec:summary}}
A phenomenological approach has been developed to quantitatively improve the $p$-$d$ elastic scattering cross section, in which additional matrix elements that serve as corrections to the 2N effective interactions are introduced.
The matrix elements were represented by a superposition of Legendre polynomials with adjustable parameters.
We performed the parameter search so that the theoretical calculation reproduces the measured differential cross sections of the $p$-$d$ elastic scattering at eight incident energies between $100$~MeV and $250$~MeV, and found the parameters with relatively smooth energy dependence.
The energy dependence was well approximated by quadratic functions, and the numerical results obtained using the parameter values estimated from the functional forms also showed good agreement with the experimental data.
It was found that the approximated energy dependence is effective at energies that were not considered in the parameter-search procedure.
These findings demonstrate that the phenomenological approach proposed in this study works well.

To investigate the 3NF contribution inside nuclei, the 2N effective interaction in free space needs to be replaced with a density-dependent 2N interaction that incorporates 3NF effects in finite nuclei.
In such calculations, the contribution from $\tilde{t}_\textrm{res}$ should be retained, as it corresponds to $p$-$d$ scattering in free space, that is, in the zero-density limit.
Therefore, the parameterization of $\tilde{t}_\textrm{res}$ presented in this paper should be regarded as a necessary step in laying the groundwork for 3NF studies in nuclei via ($p,pd$) reactions.

\section*{Acknowledgment}
The authors acknowledge H. Otsu, S. Shimoura, and K. Sekiguchi for fruitful discussions.
This work is supported by JST ERATO Grant No. JPMJER2304, Japan, and by Grants-in-Aid for Scientific Research from the JSPS (Grants No. JP21H04975, No. JP21K13919, and No. JP23KK025).

\section*{Statements and Declarations}

\noindent
{\small \textbf{Author Contributions} \
All authors contributed to the study conception and design.
Y. C. prepared the codes, carried out the calculations, and prepared the tables and figures, and the results were discussed by all authors.
Y. C. also wrote the first draft of the manuscript.
All authors commented on the draft for improvement and approved the final manuscript.}\\

\noindent
{\small \textbf{Data Availability Statement} \
The data sets generated and/or analyzed during the current study are available from the corresponding author on reasonable request.}\\

\noindent
{\small \textbf{Code Availability Statement} \
The code/software generated and/or analyzed during the current study are available from the corresponding author on reasonable request.}\\

\noindent
{\small \textbf{Conflict of Interest} \
The authors declare no conflict of interest.}

\bibliography{./References_forEPJA}

\end{document}